\documentclass [a4paper]{article}

\usepackage{fancyhdr}
\fancypagestyle{plain}{%
\fancyhead[L]{\small{DRAFT: written for Petkov (ed.), \emph{The
Ontology of Spacetime} (in preparation)}}

\addtolength{\headheight}{0.5pt} 
}

\usepackage{natbib}
\usepackage{hyperref}

\title{Minkowski space-time: a glorious non-entity}

\author{Harvey R Brown%
\thanks{Faculty of
Philosophy, University of Oxford, 10 Merton Street, Oxford OX1
4JJ, U.K.; {\em harvey.brown@philosophy.ox.ac.uk}}
\ and Oliver Pooley%
\thanks{Oriel College, Oxford OX1
4EW, U.K.; {\em oliver.pooley@philosophy.oxford.ac.uk}}}

\date{16 March, 2004}

\begin{document}

\maketitle

\begin{abstract}
It is argued that Minkowski space-time cannot serve as the deep
structure within a ``constructive'' version of the special theory
of relativity, contrary to widespread opinion in the philosophical
community.

~

This paper is dedicated to the memory of Jeeva Anandan.
\end{abstract}

\tableofcontents


\section[Einstein and the space-time explanation of
inertia]{Einstein and the space-time explanation\\ of
inertia}\label{inertia}

It was a source of satisfaction for Einstein that in developing
the general theory of relativity (GR) he was able to eradicate
what he saw as an embarrassing defect of his earlier special
theory (SR): violation of the action-reaction principle. Leibniz
held that a defining attribute of substances was their both acting
and being acted upon.  It would appear that Einstein shared this
view. He wrote in 1924 that each physical object ``influences and
in general is influenced in turn by
others.''\footnote{\citet[15]{einstein24}.}  It is ``contrary to
the mode of scientific thinking'', he wrote earlier in 1922, ``to
conceive of a thing\ldots which acts itself, but which cannot be
acted upon.''\footnote{\citet[55--6]{einstein22}. For a recent
discussion of the action-reaction principle in modern physics, see
\citet{anandanbrown95} and \citet{brown96}.}  But according to
Einstein the space-time continuum, in both Newtonian mechanics and
special relativity, is such a thing.  In these theories space-time
upholds only half of the bargain: it acts on material bodies
and/or fields, but is in no way influenced by them.

It is important to ask what kind of action Einstein thought is
involved here. Although he did not describe them in these terms,
it seems that he had in mind the roles of the four-dimensional
absolute affine connection in each case, as well as that of the
conformal structure in SR. The connection determines which paths
are geodesics, or straight, and hence determines the possible
trajectories of force-free bodies. The null cones in SR in turn
constrain the possible propagation of light.

\begin{quote}
The inertia-producing property of this ether [Newtonian
space-time], in accordance with classical mechanics, is precisely
\textit{not} to be influenced, either by the configuration of
matter, or by anything else. For this reason, one may call it
``absolute''.  That something real has to be conceived as the
cause for the preference of an inertial system over a noninertial
system is a fact that physicists have only come to understand in
recent years ... Also, following the special theory of relativity,
the ether was absolute, because its influence on inertia and light
propagation was thought to be independent of physical influences
of any kind ... The ether of the general theory of relativity
differs from that of classical mechanics or the special theory of
relativity respectively, insofar as it it is not ``absolute'', but
is determined in its locally variable properties by ponderable
matter. \citep{einstein24}
\end{quote}

The success in salvaging the action-reaction principle was not
confined in GR to the fact that the space-time metric field (which
of course determines both the connection, by the principle of
metric compatibility, and the conformal structure) is dynamical,
being a solution to Einstein's field equations, which couple
matter degrees of freedom to the metric. In the early twenties,
when he wrote the above comment, Einstein had still not discovered
an important aspect of his theory of gravitation---the fact that
the field equations themselves underpin the geodesic principle.
This principle states that the world-lines of force-free test
particles are constrained to lie on geodesics of the connection.
It is important for our purposes to dwell briefly on the
significance of this fact.

We have seen that for Einstein the inertial property of
matter\footnote{To be precise, the fact that particles with
non-zero mass satisfy Newton's first law of motion, not that they
possess such inertial mass} requires explanation in terms of the
action of a real entity on the particles. It is the space-time
connection that plays this role: the affine geodesics form ruts or
grooves in space-time that guide the free particles along their
way. The intuition was well expressed by Nerlich in 1976:
\begin{quote}
\ldots without the affine structure there is nothing to determine
how the [free] particle trajectory should lie. It has no antennae
to tell it where other objects are, even if there were other
objects\ldots It \textit{is because space-time has a certain shape
that world lines lie as they do}. \citep[264, original
emphasis]{Nerlich76}
\end{quote}

In GR, on the other hand, this view is at best redundant, at worst
problematic. For it follows from the form of Einstein's field
equations that the covariant divergence of the stress-energy
tensor field $T_{\mu\nu}$---that object which incorporates the
``matter'' degrees of freedom---vanishes.

\begin{equation}
T^\mu{}_{\nu;\mu} = 0
\end{equation}
This result is about as close as anything is in GR to the
statement of a conservation principle, and it came to be
recognised as the basis of a proof, or proofs, that the
world-lines of a suitably modelled force-free test particles are
geodesics.\footnote{See, for example, \citet[\S20.6,
471--80]{mtw73}.} The fact that these proofs vary considerably in
detail need not detain us. The first salient point is that the
geodesic principle for free particles is no longer a postulate but
a theorem. GR is the first in the long line of dynamical theories,
starting with the Aristotelian system and based on that profound
distinction between natural and forced motions of bodies, that
\textit{explains} inertial motion. The second point is that the
derivations of the geodesic principle in GR also demonstrate its
limited validity. In particular, it is not enough that the test
particle be force-free. It has long been recognised that spinning
bodies for which tidal gravitational forces act on its elementary
pieces deviate from geodesic behaviour.\footnote{See \citet[480;
ex.~40.8, 1120--1; and \S40.9, 1126--31]{mtw73}. These authors
refer briefly on p.~480 to the complications that quantum physics
is likely to introduce to the question of geodesic behaviour. We
note that the familiar picture of light tracing out the null cones
of space-time is also probably only approximately valid (though
the approximation is usually extremely good) as a result of
quantum physics.  Since 1980, studies have been made of the
propagation of photons in QED in curved space-times, in the
esoteric regime where the scale of the space-time curvature is
comparable to the Compton wavelength of the electron. Here, vacuum
polarisation causes the vacuum to act both as a dispersive and
birefringent optical medium. In particular the propagation of
photons as determined by geometric optics is controlled by an
effective metric that differs from the space-time metric
$g_{\mu\nu}$. For a recent review paper, see \citet{shore03}.}
What this fact should clarify, if indeed clarification is needed,
is that it is not simply \emph{in the nature} of force-free bodies
to move in a fashion consistent with the geodesic principle. It is
not an essential property of localised bodies that they run along
the ruts of space-time determined by the affine connection, when
no other dynamical influences are at play. In Newtonian mechanics
and SR, the conspiracy of inertia is a postulate, and its putative
explanation by way of the affine connection is a postulate added
to a postulate.

And it is here that Einstein and Nerlich part company with
Leibniz, and even Newton. For both Leibniz \textit{and Newton},
absolute space-time structure is not the sort of thing that acts
at all. If this is correct, and we believe it is, then neither
Newtonian mechanics nor SR represent, \textit{pace} Einstein, a
violation of the action-reaction principle, because the space-time
structures in both cases are neither acting nor being acted upon.
Indeed we go further and agree with Leibniz that they are not real
entities in their own right at all.

\section{The nature of absolute space-time}

It is well known that Leibniz rejected the reality of absolute
Newtonian space and time principally on the grounds that their
existence would clash with his principles of Sufficient Reason and
the Identity of Indiscernibles. Nonentities do not act, so for
Leibniz space and time can play no role in explaining the mystery
of inertia.

Newton seems to have agreed with this conclusion, but for
radically different reasons, as expressed in his
pre-\textit{Principia} manuscript \textit{De Gravitatione}
\citep{newton62}. For Newton, the existence of absolute space and
time has to do with providing a structure, necessarily distinct
from ponderable bodies and their relations, with respect to which
it is possible systematically to define the basic
\textit{kinematical} properties of the motion of such bodies. For
Newton, space and time are not substances in the sense that they
can act, but are real things nonetheless.\footnote{It is worth
stressing that its lack of causal influence is Newton's sole
reason for refraining from calling space a substance.  It is
therefore at least misleading to deny that Newton was a
substantivalist.} It is now known, however, that the job can be
done without postulating any background space-time scaffolding,
and that at least a significant subset---perhaps \textit{the}
significant subset---of solutions to any Newtonian theory can be
recovered in the process.\footnote{The discovery was made by
Julian Barbour and Bruno Bertotti \citep{bb82,barbour94}.  For
discussion see \citet{belot00} and \citet{pooleybrown02}.}

Recall Nerlich's remark above to the effect that force-free
particles have no antennae, that they are unaware of the existence
of other particles. That \textit{is} the \textit{prima facie}
mystery of inertia in pre-GR theories: how do all the free
particles in the world know how to behave in a mutually
coordinated way such that their motion appears extremely simple
from the point of view of a family of privileged frames? But to
appeal to the action of a background space-time connection in
which the particles are immersed---to what Weyl called the
``guiding field''---is arguably to enhance the mystery, not to
remove it. For the particles do not have space-time feelers
either.  In what sense is the postulation of the 4-connection
doing more explanatory work than Moliere's famous dormative virtue
in opium? (We return to this question below.)

It is of course non-trivial that inertia can be given a
\textit{geometrical} description, and this is associated with the
fact that the behaviour of force-free bodies does not depend on
their constitution: it is universal. But again what is at issue is
the arrow of explanation. In our view it is simply more economical
to consider the 4-connection as a codification of certain key
aspects of the behaviour of particles and fields.\footnote{One
faces a similar choice in parity-violating theories: do
orientation fields play an explanatory role in such theories, or
are they simply codifications of the coordinated asymmetries
exhibited by the solutions of such theories?  See
\citet[272--4]{pooley03}.}

\section{The principle \textit{vs}.\ constructive theory distinction}

In recent years there has been increasing discussion of the role
that thermodynamics played as a methodological template in
Einstein's development of special relativity, and of his
characterization of SR as a ``principle'' theory, as opposed to a
``constructive'' theory like the kinetic theory of
gases.\footnote{An excellent characterisation of the
principle-constructive distinction is found in \citet[331]{bj}. We
have much to say about their paper in what follows.  For other
recent discussions of the role played by the distinction in the
history and philosophy of SR, see \cite{brownpooley01} and
\cite{brown03}.}

The distinction is not a categorical one, nor must a principle
theory be bereft of any constructive elements. What we have
effectively argued in section~\ref{inertia} is that Einstein's
comments in the 1920s on the role of the Newtonian and SR
``ethers'', or space-times, indicate that he came to interpret
inertial structure as a genuinely constructive element in these
theories.  (In our view it is unwarranted to attribute the same
view to Einstein around 1905.) However, relativistic effects such
as length contraction and time dilation are another matter. It is
clear that in 1905, and for many subsequent years, Einstein
regarded their derivation in SR as akin to the derivation in
thermodynamics of, say, the existence of entropy as a
thermodynamic coordinate---as being, that is to say, a necessary
condition for the validity of certain phenomenological principles
that themselves have only empirical robustness as their
justification.\footnote{\label{sommletter}It is widely known that
the fullest account given by Einstein of the claim that SR has the
nature of a `principle-theory' was in an article on relativity
theory he was commissioned to write in 1919 for The Times of
London \citep{einstein19}.  Should it be thought that the popular
nature of the publication and/or its date lessen the degree to
which Einstein's claim is to be taken seriously, two points might
be borne in mind. First, the claim is entirely consistent with the
story of Einstein's pre-1905 struggles with the constructive
approach to electrodynamics and the theory of the electron---which
were based largely on the difficulties posed by the emergence of
Planck's constant (see below, pp.~\pageref{plancknotrel}ff).
Secondly, the methodological analogy between SR and thermodynamics
was mentioned by Einstein on several occasions prior to 1919. In a
short paper of 1907 replying to a query of Ehrenfest on the
deformable electron, he wrote:
\begin{quote}The principle of relativity, or, more exactly, the
principle of relativity together with the principle of the
constancy of velocity of light, is not to be conceived as a
``complete system'', in fact, not as a system at all, but merely
as a heuristic principle which, when considered by itself,
contains only statements about rigid bodies, clocks, and light
signals. It is only by requiring relations between otherwise
seemingly unrelated laws that the theory of relativity provides
additional statements. \dots we are not dealing here at all with a
``system'' in which the individual laws are implicitly contained
and from which they can be found by deduction alone, but only with
a principle that (similar to the second law of the theory of heat)
permits the reduction of certain laws to others.
\citep{einstein07}\end{quote} In a letter to Sommerfeld of 1908,
Einstein wrote:
\begin{quote}The theory of relativity is not more conclusively and
absolutely satisfactory than, for example, classical
thermodynamics was before Boltzmann had interpreted entropy as
probability. If the Michelson-Morley experiment had not put us in
the worst predicament, no one would have perceived the relativity
theory as a (half) salvation. Besides, I believe that we are still
far from having satisfactory elementary foundations for electrical
and mechanical processes.\citep[50]{einstein08}\end{quote}}

We have discussed elsewhere Einstein's recognition of the fact
that constructive theories have more explanatory power than
principle theories, as well as the misgivings that he expressed,
particularly late in his life, about the appropriateness of his
separation of kinematical and dynamical considerations in the 1905
paper \citep{brownpooley01}. What we wish to consider here is the
question of the possibility of a fully constructive rendition of
SR, and in particular the possibility of a constructive
explanation of the `kinematical' effects associated with length
contraction and time dilation.

The issues surrounding this question have been discussed recently
by \citet{bj}.  As will soon become clear, we take a different
view to them about what might constitute a constructive version of
SR.  However, before addressing this issue directly, we want to
return briefly to the claim that principle theories lack the
explanatory power of constructive theories, for this, too, is an
issue addressed by Balashov and Janssen.

Balashov and Janssen see no problem with the idea that Einstein's
original principle-theory presentation of SR can be held to
explain the phenomenon of length contraction. They write:
``Understood purely as a theory of principle, SR explains this
phenomenon if it can be shown that the phenomenon necessarily
occurs in any world that is in accordance with the relativity
postulate and the light postulate.''  They concede that, in
contrast to constructive-theory explanations, such a
principle-theory explanation will ``have nothing to say about the
reality behind the phenomenon''  \citeyearpar[331]{bj}.

Later in their paper, which is a critical review of aspects of
William Lane Craig's recent writings in defence of presentism
\citep{craig00a,craig00b,craig01}, they take explicit issue with
two claims that they attribute to Craig: (i) that SR in its 1905
form fails to provide a theory-of-principle explanation of
phenomena such as length contraction and, (ii) that
theory-of-principle explanations in general are deficient
\citeyearpar[332]{bj}.  We side with Craig on both counts,
although it should be stressed that we endorse (i) for reasons
quite different to those that motivate Craig.  Before outlining
our reasons for rejecting the idea that Einstein's 1905 derivation
of the Lorentz transformations can provide any sort of explanation
of length contraction we mention Balashov and Janssen's main
reason for contesting (ii).\footnote{They also note that
principle-theory derivations of particular phenomena fit the
covering law model of explanation, but, as they rightly concede,
this ``might just be another nail in the coffin of the covering
law model'' \cite[332]{bj}.}

It rests, simply, in their noting that (ii) applies equally to
thermodynamics: ``That in and of itself, we submit, places the
relativity interpretation [i.e., Einstein's 1905 presentation of
SR] in very good company'' \citeyearpar[332]{bj}. It is certainly
true that Einstein's original derivation of SR is in good company,
but this company is not necessarily a company rich in explanatory
resources. Balashov and Janssen are prepared to admit that
Einstein thought that principle theories were ``inferior'' to
constructive theories, but this rather general claim might seem to
miss the very point of Einstein's articulation of the constructive
versus principle theory distinction, and his citation of
thermodynamics as a paradigm example of a principle theory.
Einstein's view (one that we share) was that principle theories
were `inferior' specifically in their explanatory power.  His
contrasting thermodynamics, as a principle theory, with
statistical mechanics, as a constructive theory, was supposed to
illustrate precisely that:
\begin{quotation}
It seems to me \ldots that a physical theory can be satisfactory
only when it builds up its structures from \textit{elementary}
foundations. \citep{einstein08}\footnote{It is clear to us that in
his 1908 letter, by the term ``elementary foundations'' Einstein
means the building blocks of a constructive theory (see
footnote~\ref{sommletter} above, which contains the sentences from
the letter that follow on immediately from the sentence just
quoted). However, Stachel appears to think that the term refers to
principles akin to those of thermodynamics
\citep[xxii]{einstein90}.}

\ldots when we say we have succeeded in understanding a group
  of natural processes, we invariably mean that a constructive theory
  has been found which covers the processes in question. \cite[228]{einstein82}
\end{quotation}
It is certainly not the case that Einstein viewed principle
theories as inferior in other respects.  As Balashov and Janssen
rightly note, their founding principles often enjoy particularly
strong empirical confirmation.  Einstein is well known for having
had greater confidence in the laws of thermodynamics (and, for the
same reason, in SR) than in any other laws of physics.

An examination of the status of length contraction in the context
of Einstein's 1905 treatment of SR will illustrate the way in
which principle theories fail to be explanatory.  Recall that in
this derivation the first conclusion drawn from the two
fundamental postulates is the invariance of the speed of light,
that it has the same constant value in all inertial frames.  This
gives the `$k$-Lorentz transformations', the Lorentz
transformations up to a velocity dependent scale factor, $k$. What
has, in effect, been shown is that if the speed of light as
measured with respect to frame $F'$ is to be found to be the same
value as when measured with respect to the `resting frame' $F$,
then rods and clocks at rest in $F'$ had better contract and
dilate (with respect to frame $F$) in the coordinated way that is
encoded in the $k$-Lorentz transformations.  One then appeals to
the relativity principle again---the principle entails that these
coordinated contractions and dilations must be exactly the same
function of velocity for each inertial frame---, along with the
principle of spatial isotropy, in order to narrow down the
deformations to just those encoded in the Lorentz
transformations.\footnote{The importance of this second
application of the relativity principle (together with spatial
isotropy) in Einstein's 1905 logic was stressed in \citet{brown97}
and particularly in \citet{brownpooley01}. Janssen points out that
in 1905 Einstein ``found part of another result that had been
found by Poincar\'e, namely, that the Lorentz transformations form
what mathematicians call a \emph{group}'' \cite[428]{janssen01}.
But it is important to realize that it is appeal to the relativity
principle that justifies the fact that the coordinate
transformations form a group. The group property is essentially a
postulate in Einstein's reasoning, not a theorem.} What has been
shown is that rods and clocks must behave in quite particular ways
in order for the two postulates to be true together.  But this
hardly amounts to an explanation of such behaviour.  Rather things
go the other way around.  It is \emph{because} rods and clocks
behave as they do, in a way that is consistent with the relativity
principle, that light is measured to have the same speed in each
inertial frame.

We now return to the question of what might constitute a
constructive version of SR.  It is useful in this connection to
start by recalling that Einstein had not adopted the principle
theory route to SR by chance.\label{plancknotrel} He was familiar
with Lorentz's semi-constructive efforts in the 1890s to account
for the null result of the 1887 Michelson-Morley experiment in
terms of a postulated shape deformation suffered by solid bodies
as a result of motion through the luminiferous ether. But by 1905
Einstein had convinced himself for a number of reasons that a
systematic understanding of the non-Newtonian behaviour of moving
rods and clocks based on the study of the forces holding their
constituent parts together was, at that time, far too ambitious.

Note that Einstein did not reject the approach initiated by
Lorentz primarily because it violated the relativity principle.
Although Lorentz believed in a preferred inertial frame, by 1904
the kinematics of his theory of the electron were consistent with
the relativity principle.  His theorem of corresponding states was
based on the assumption that no experiment could be performed that
would exhibit the presence of the ether, at least as regards
effects that were up to second order in $v/c$; for all predictive
purposes Lorentz's theory of the electron had become compatible
with the relativity principle. What instead concerned Einstein was
the confused state of understanding---exacerbated by his own
revolutionary hypothesis of light quanta!---of the stability of
matter in terms of the dynamical forces operating at the atomic
and molecular levels.

By the late 1940s, a much better picture, at least in broad terms,
of the cohesion of matter was available.  Even so, in his 1949
\textit{Autobiographical Notes}\footnote{See \citet{einstein69}.}
Einstein's reservations about quantum mechanics apparently
prevented his re-examining the constructive route to SR, despite
his now articulating clear misgivings about key aspects of his
1905 principle theory approach.  But what is especially striking
is this.  We saw above in section~\ref{inertia} that in 1922
Einstein referred to the SR ``ether'' as having an ``influence on
light propagation,'' but in the 1949 \textit{Notes} he warns
against imagining that space-time intervals ``are physical
entities of a special type, intrinsically different from other
variables (`reducing physics to geometry', etc.).''

Since the twenties there has been a small minority of
voices---including those of Pauli, Eddington, Swann and
Bell---defending, to a greater or lesser extent, the importance of
a \emph{constructive}, non-geometric picture of the kinematics of
SR that makes no commitment to the existence of a preferred
inertial frame. We have added our voices this little-known
tradition \citep[see][]{brown93, brown97,brownpooley01,brown03}.
Recently we dubbed the approach, following a remark of John Bell
\citeyearpar[77]{bell76}, the ``Lorentzian pedagogy''. This label
has proved to have several unfortunate and highly misleading
features, the worst being that any position named after Lorentz
risks being misinterpreted as an endorsement of a preferred
frame.\footnote{The remaining unfortunate features are these.
First, the pedagogic dimension offered by Bell's simple atomic
model displaying motion-induced relativistic contraction is not
the whole story, as Bell himself recognized \citep[see
also][]{brownpooley01}. Secondly, as has been argued recently in
\citet{brown03}, the term ``FitzGeraldian pedagogy'' would be
historically more appropriate. Finally, in so far as there is a
connection with Lorentz's thinking, it is only his post-1905
formulation of the electron theory, in which Lorentz had learnt
from Einstein how correctly to interpret the Lorentz
transformations \citep[see][8]{janssen01} that is relevant---but
shorn of the privileged frame!} A more appropriate label would be
the \emph{dynamical interpretation}.\footnote{Such an approach
does not appear (under any label) within the recent taxonomy of
interpretations of SR produced by Craig, and endorsed, with
qualification, by Balashov and Janssen.}

But as one of us has noted \citep{brown97} the `space-time theory'
approach developed principally by philosophers in North America in
recent decades---the view of SR that is encapsulated in Friedman's
1983 book \textit{Foundations of Space-Time Theories}---also could
be interpreted as a constructive theory in Einstein's sense, where
it is precisely the Minkowski geometry that provides the
explanatory deep structure. An explicit defence of this position
has recently been given by Balashov and Janssen, to which we now
turn.

\section{The explanation of length contraction}

How \emph{are} we to explain length contraction in SR?  One needs
to be careful about what, exactly, is taken to stand in need of an
explanation.

Balashov and Janssen's initial characterization of the
constructive-theory explanation of the space-time interpretation
runs as follows:
\begin{quote}
length contraction is explained by showing that two observers who
are in relative motion to one another and therefore use different
sets of space-time axes disagree about which cross-sections of the
`world-tube' of a physical system give the length of the system.
\citeyearpar[331]{bj}
\end{quote}
Here we are asked to contemplate a single rod.  What is to be
explained is how it is possible that this single rod comes to be
assigned two different lengths when measured with respect to two
inertial frames.  Note that the relativity of simultaneity---that
two different cross-sections of the rod are involved---plays a
crucial role.\footnote{In a recent manuscript, Petkov claims to
show that ``no forces are involved in the explanation of the
Lorentz contraction'' \citep[6]{petkov02}.  His argument involves
consideration of essentially the same scenario considered by
Balashov and Janssen.  And, of course, those who believe (like us)
that in some explanatory contexts it is correct to invoke forces
would not do so when comparing one cross-section of the world tube
of a rod with another cross-section of the same rod. Rather,
forces are relevant, for example, when comparing, \emph{relative
to a fixed inertial frame and standard of simultaneity}, two
otherwise identical rods that are in different states of motion
relative to this frame of reference.}

In his contribution to this volume \cite[xxx]{saunders04},
Saunders considers two rods, $R$ and $S$, in relative inertial
motion. Specific features of Minkowski geometry are appealed to in
an explanation of why, \emph{relative to surfaces of simultaneity
orthogonal to the world-tube of $R$}, $S$ is shorter than $R$
whereas, \emph{relative to surfaces of simultaneity orthogonal to
the world-tube of $S$}, it is $R$ that is shorter than
$S$.\footnote{An analogous scenario is also considered by
\citet[499--500]{janssen02}.}

In our opinion these constitute perfectly acceptable explanations
(perhaps the only acceptable explanations) of the explananda in
question.  But it is far from clear that they qualify as
\emph{constructive} explanations.\footnote{It should be stressed
that Saunders does not claim that the explanation he sketches is a
constructive-theory explanation.} What is being \emph{assumed} in
both cases is that the rod(s) being measured, and the rods and
clocks doing the measuring, all satisfy the constraints of
Minkowskian geometry. The explanations point out that if objects
obeying these constraints have certain geometrical features, then
it follows, as a simple consequence of the mathematics of
Minkowskian geometry, that they will have certain other features.

The geometrical features of the objects that are assumed, and
appealed to, in these explanations are similar in status to the
postulates of principle theories.  They do not, \emph{directly},
concern the details of the bodies' microphysical constitution.
 Rather they are about aspects of their (fairly) directly observable
macroscopic behaviour. And this reflection prompts an obvious
question: \emph{why} do these objects obey the constraints of
Minkowski geometry?\footnote{Note that this question arises for
someone with no prior expectations about how bodies in motion
`should' behave; \emph{pace} \citet[340]{bj}, the question need
not
 be understood as asking ``why do these objects obey the
constraints of Minkowski geometry \emph{rather than those of
Newtonian space-time}?''} It is precisely this question that calls
out for a constructive explanation. What sort of an answer might
be given?

The following quote from Friedman helps to delineate the options.
In discussing Poincar\'{e}'s preference for ``the
Lorentz-Fitzgerald version of an `aether' theory'' over Einstein's
formulation of SR he writes:
\begin{quote}
\ldots the crucial difference between the two theories, of course,
is that the Lorentz contraction, in the former theory, is viewed
as a result of the (electromagnetic) forces responsible for the
microstructure of matter in the context of Lorentz's theory of the
electron, whereas this same contraction, in Einstein's theory, is
viewed as a direct reflection---independent of all hypotheses
concerning microstructure and its dynamics---of a new kinematical
structure for space and time involving essential relativized
notions of duration, length, and simultaneity.  In terms of
Poincar\'{e}'s hierarchical conception of the sciences, then,
Poincar\'{e} locates the Lorentz contraction (and the Lorentz
group more generally) at the level of experimental physics, while
keeping Newtonian structure at the next higher level (what
Poincar\'{e} calls mechanics) completely intact.  Einstein, by
contrast, locates the Lorentz contraction (and the Lorentz group
more generally) at precisely this next higher level, while
postponing to the future all further discussion of the physical
forces and material structures actually responsible for the
physical phenomenon of rigidity.  The Lorentz contraction, in
Einstein's hands, now receives a direct \emph{kinematical}
interpretation. \cite[211--2]{friedman02}
\end{quote}

The talk of a preference for one theory over the the other might
suggest that we are dealing with two incompatible, rival
viewpoints. On one side one has a truly constructive space-time
interpretation of SR, involving the postulation of the structure
of Minkowski space-time as an ontologically autonomous element in
the models of the phenomena in question. In this picture, length
contraction is to be given a constructive explanation in terms of
Minkowski space-time because complex material bodies are
constrained (somehow!) to ``directly reflect'' its structure, in a
way that is ``independent of all hypotheses concerning
microstructure and its dynamics.''\footnote{The thesis that
Minkowski spacetime cannot act in this way as an
\textit{explanans} in a constructive version of SR was put forward
in \cite{brown97} and further defended in \cite{brownpooley01} and
\cite{brown03}.} If one were to adopt such a viewpoint there would
seem to be little room left for the alternative viewpoint,
according to which the explanation of length contraction is
ultimately to be sought in terms of the 
dynamics of
the microstructure of the contracting rod.

In fact, it is not clear that Friedman has these two opposing
pictures in mind.  Although he claims that \emph{Poincar\'{e}}
keeps Newtonian structure at the level of `mechanics', if one is
committed to the idea that Lorentz contraction is the result of
the forces responsible for the microstructure of matter then one
should, in our opinion, believe that Minkowskian, rather than
Newtonian, structure is the appropriate kinematics for mechanics.
In our view, the appropriate structure is Minkowski geometry
\emph{precisely because} 
the laws of physics, including those to be appealed to in the
dynamical explanation of length contraction, are Lorentz
covariant. Equally one can postpone (as Einstein did) the detailed
investigation into the forces and structures actually responsible
for the phenomena that are paradigmatic of space-time's
Minkowskian geometry without thereby relinquishing the idea that
these forces and structures are, indeed, ``actually responsible''
for the phenomena in question and, hence, (we go further in
suggesting) for space-time having the structure that it has.

Saunders is critical of the Lorentzian pedagogy because he takes
it to \emph{require} that the investigation of dynamical phenomena
is to be referred to a single (though arbitrary) frame of
reference. It is true that Bell was concerned to extol the virtues
of working wholly within a single frame.  But on this score his
point was primarily a pedagogic one.  His point was not that one
is required to work in a single framed, but that one always
\emph{can} work in a single frame.  In particular, he was
concerned to show that, if one knew the laws of physics with
respect to a given frame one could, at least in principle,
\emph{derive} how they should be described with respect to other
frames.\footnote{If the laws known with respect to a given frame
are in fact (though not known to be) Lorentz covariant, one will
derive that the rods and clocks at rest in another frame will be
contracted and dilated relative to one's own: one will derive that
a Lorentz transformation is the correct coordinate transformation
relating the two frames.  One can then go on to investigate how
phenomena in general are to be described relative to this frame,
and to derive that these descriptions will obey laws of exactly
the same form as do descriptions with respect to one's own frame.
One will have thereby derived the Lorentz covariance of the laws
\citetext{see \citealp[75--6]{bell87}; \emph{cf}.\
\citealp{swann41}}.}  Bell believed that exploiting the
perspective one gains from working with respect to a single frame
best allows one to discern the great continuity that exists
between relativity and the physics that predated Einstein's 1905
paper. As such, the single-frame perspective is a useful antidote
to misapprehensions about relativity that arise when one focusses
solely on the discontinuities.

But focus on describing all phenomena with respect to a single
frame is just one part of Bell's message.  Moreover, it is not
that part which forms the essential element in the position we
have called the dynamical interpretation. What is definitive of
this position is the idea that constructive explanation of
`kinematic' phenomena involves investigation of the details of the
dynamics of the complex bodies that exemplify the kinematics.

And it seems that Saunders agrees on this score.  Given a
word-line that represents the possible trajectory of the end point
of a small rod, and given a single point that is meant to
represent the other end of the rod at some particular moment,
there is, from the point of view of Minkowski geometry, a
particularly natural construction of a second curve through the
single point. The two curves together define the possible
world-tube of an extended body, a world-tube that, from the point
of view of Minkowski geometry, is particularly natural. According
to Saunders ``it is this construction that needs a dynamical
underpinning: why do stable bodies, sufficiently small in size,
have world-tubes with this geometry?'' \cite[xxx]{saunders04}.
This, we claim, is precisely the type of question that the
dynamical interpretation of SR seeks to address.  Little hangs on
whether the dynamical underpinning is spelled out with respect to
a particular frame, or whether the solution is given in some
sophisticated, coordinate independent way.  What is important is
that particular laws---a specific quantum field theory---could be
solved and the solutions shown to have the requisite geometrical
properties.\footnote{It is perhaps worth mentioning here one
common objection to any approach that seeks to reduce
non-dynamical space-time structure, such as that of SR, to the
symmetries of the laws governing matter. According to the
objection such an approach is constrained to use special
coordinates (in which the laws take their canonical form) because
otherwise the geometric structure, in the form of the Minkowski
metric and its connection coefficients, appears explicitly in the
laws.

Two things should be said in response to this objection. First,
the objection surely is not that defenders of the reductive
account are okay so long as they place restrictions on admissible
coordinate systems.  If the reductive account is tenable at all,
then it can countenance the use of arbitrary coordinate systems.
(Perhaps it could be argued that the supporter of the reductive
account faces an obligation to provide an alternative formulation
of the laws as written with respect to arbitrary coordinates, in
which the secondary status of the geometrical structure is clear.)
Secondly, even if one is inclined to take the appearance of the
Minkowski metric in the laws when written generally covariantly as
a reason to afford it a primitive ontological status, one is still
obliged to tell some story about how and why material systems
reflect its structure in their macroscopic behaviour.  What could
this story be, other than the dynamical one?  (In a recent
discussion, Butterfield also recognises the existence of such an
obligation, which, in his terminology, is an obligation to answer
to the ``consistency problem'' \citep[\S2.1.2]{butterfield01}.)}

We have been arguing that the truly constructive explanation of
length contraction involves solving the dynamics governing the
structure of the complex material body that undergoes contraction.
There are, of course, many contexts in which such an explanation
may not be appropriate, contexts that call for a purely
geometrical explanation. What we wish to stress is (i) that such
geometrical explanations are not constructive theory explanations
in Einstein's sense and (ii) that there \emph{are} contexts, and
questions, to which the dynamical story is appropriate.

There is one final, important, area of disagreement between
ourselves and Balashov and Janssen to map out.  But before we do
so, it will be instructive to acknowledge that in many contexts,
perhaps in most contexts, one should not appeal to the
\emph{details} of the dynamics governing the microstructure of
bodies exemplifying relativistic effects when one is giving a
constructive explanation of them.\footnote{We thank Michel Janssen
for reminding us of this point.} \emph{Granted that there are
stable bodies}, it is sufficient for these bodies to undergo
Lorentz contraction that the laws (whatever they are) that govern
the behaviour of their microphysical constituents are Lorentz
covariant. It is \emph{the fact that the laws are Lorentz
covariant}, one might say, that explains why the bodies Lorentz
contract. To appeal to any further details of the laws that govern
the cohesion of these bodies would be a mistake.

Elsewhere we have dubbed this view the ``truncated'' Lorentzian
pedagogy \citep[261]{brownpooley01}.  It is worth making two
points about it.  First, to explain why there are any bodies at
all that conform to Minkowskian geometry one needs to appeal to
more than Lorentz covariance.  One needs to demonstrate the
possibility of stable material configurations, and the
constructive explanation of this will involve a more complete
dynamical analysis. Secondly, one might be tempted to deny that
explanations which appeal to an explanans as non-concrete as the
\emph{symmetries} of the laws are genuinely constructive
explanations. In other words, it turns out that there are even
fewer contexts than one might have at first supposed in which
length contraction stands in need of a constructive-theory
explanation.

\section{Minkowski space-time: the cart or the horse?}

But if it is often sufficient to appeal to Lorentz covariance to
give a dynamical explanation of length contraction, is that where
explanations should stop?  It is here that Balashov and Janssen
see a further, constructive role for the geometry of space-time.
They ask:
\begin{quote}
... does the Minkowskian nature of space-time explain why the
forces holding a rod together are Lorentz invariant or the other
way around? Our intuition is that the geometrical structure of
space(-time) is the \emph{explanans} here and the invariance of
the forces the \emph{explanandum}. To switch things around, our
intuition tells us, is putting the cart before the horse.
\citep[340--1]{bj}
\end{quote}
The same issue was raised some years ago in \citet{brown93} and,
particularly, \cite{brown97}, but there the opposite view to
Balashov and Janssen's was taken as to what was to be regarded as
the cart and what the horse.

It is worth recalling that Balashov and Janssen's target is the
particular neo-Lorentzian interpretation of SR advocated by Craig.
This is an interpretation in which space-time structure is
supposed to be Newtonian and in which there is supposed to be a
preferred frame, consistent with Craig's commitment to a tensed
theory of time.  Balashov and Janssen's claim is that the
space-time interpretation has a definite explanatory advantage
over this neo-Lorentzian interpretation when it comes to the
Lorentz covariance of the laws governing the behaviour of matter:
\begin{quotation}
In the former, Lorentz invariance reflects the structure of the
space-time posited by the theory. In the latter, Lorentz
invariance is a property accidentally shared by all laws
effectively governing systems in Newtonian space and time\ldots

In the neo-Lorentzian interpretation it is, in the final analysis,
an unexplained coincidence that the laws effectively governing
different sorts of matter all share the property of Lorentz
invariance, which originally appeared to be nothing but a
peculiarity of the laws governing electromagnetic fields. In the
space-time interpretation this coincidence is explained by tracing
the Lorentz covariance of all these different laws to a common
origin: the space-time structure posited in this interpretation
(Janssen [1995], [2002])\ldots No matter how the argument is made,
the point is that there are brute facts in the neo-Lorentzian
interpretation that are explained in the space-time
interpretation. As Craig (p. 101) writes (in a different context):
`if what is simply a brute fact in one theory can be given an
explanation in another theory, then we have an increase in
intelligibility that counts in favor of the second theory.'
\end{quotation}

We agree that in Craig's neo-Lorentzian interpretation of SR, and
according to our preferred dynamical interpretation, the Lorentz
covariance of all the fundamental laws of physics is an
unexplained brute fact.  This, in and of itself, does not count
against the interpretations: all explanation must stop somewhere.
What is required if the so-called space-time interpretation is to
win out over the dynamical interpretation (and Craig's
neo-Lorentzian interpretation) is that it offers a genuine
explanation of Lorentz covariance.  This is what we dispute.  Talk
of Lorentz covariance ``reflecting the structure of space-time
posited by the theory'' and of ``tracing the invariance to a
common origin'' needs to be fleshed out if we are to be given a
genuine explanation here---something akin to the explanation of
inertia in general relativity (see section 1 above). Otherwise we
simply have yet another analogue of Moliere's dormative virtue.

In fact Balashov and Janssen's own example can be turned against
them.  Craig's neo-Lorentzian interpretation is precisely an
example of theory in which the symmetries of spacetime structure
are not reflected in the symmetries of the laws governing matter.
Balashov and Janssen do not question the coherence of this theory
(as we would). Rather they seek to rule it out on the grounds of
its explanatory deficiencies when compared to their preferred
theory. This shows that, as matter of logic alone, if one
postulates spacetime structure as a self-standing, autonomous
element in one's theory, it need have no constraining role on the
form of the laws governing the rest of content of the theory's
models.\footnote{See in this connection \citet{brown93}. It might
be useful to recall here the example of the approach to Einstein's
field equations in GR based on the the introduction of a spin-2
field on flat Minkowski space-time \citep[for references,
see][xiii--iv]{preskillthorne99}. The operational significance of
the background space-time in this theory is not the same as that
in SR.} So how is its influence on these laws supposed to work?
Unless this question is answered, space-time's Minkowskian
structure cannot be taken to explain the Lorentz covariance of the
dynamical laws.

From our perspective, of course, the direction of explanation goes
the other way around.  It is the Lorentz covariance of the laws
that underwrites the fact that the geometry of space-time is
Minkowskian.  It is for this reason that we can rule out the sort
of mismatch between space-time symmetries and dynamical symmetries
that are a feature of Craig's interpretation, and that so trouble
Balashov and Janssen.

Balashov and Janssen acknowledge that some of their readers will
have this `relationist' intuition.  Remarkably they claim that
this does not weaken their point!  In a footnote, they admit that,
for the relationalist, the Lorentz covariance of the laws ``in a
sense does seem to explain'' why space-time structure in
Minkowskian.  (We, of course, see no reason for their
qualifications here.)  But, they go on to assert that the
relationalist should nonetheless view, for example, the Euclidean
nature of space as explaining why the forces holding Cyrano's nose
together are invariant under rotations rather than \emph{vice
versa} \citep[341, fn~11; \emph{cf.}\ 340]{bj}.  As far as we can
see, this amounts to bald assertion.  We happily concede that
there are many contexts in which the Euclidean nature of space is
the appropriate explanation of the behaviour of Cyrano's nose. But
we insist that there are others in which it is appropriate to
appeal to the Euclidean symmetries of the forces at work to
explain the same behaviour.  And we simply deny that the Euclidean
nature of space can ever be cited as a genuine explanation of
these symmetries; \emph{this} would be to put the cart before the
horse.

A more sustained discussion of Minkowski space-time's providing a
putative common origin for the ``unexplained coincidence'' in
Lorentz's theory that both matter and fields are governed by
Lorentz covariant laws, is to be found in Janssen's detailed
recent analysis of the differences between the Einstein and
Lorentz programs \citep{janssen01}.  It is also covered in his
wider investigation of `common origin inferences' in the history
of science \cite[497--507]{janssen02}.  In our view, neither of
these papers succeed in clarifying how space-time structure can
act as a ``common origin'' of otherwise unexplained coincidences.
One might, for example, go so far as to agree that all particular
instances of paradigmatically relativistic kinematic behaviour are
traceable to a common origin: the Lorentz covariance of the laws
of physics.  But Janssen wants us to go further.  He wants us to
then ask after the common origin of this universal Lorentz
covariance.  It is his claim that this can be traced to the
space-time structure posited by Minkowski that is never clarified.

For example, immediately after making this claim in
\citet{janssen02}, he writes:
\begin{quote}
In Minkowski space-time, the spatio-temporal coordinates of
different observers are related by Lorentz transformations rather
than Galilean transformations.  Any laws for systems in Minkowski
space-time must accordingly be Lorentz invariant.
\end{quote}
There is an dangerous ambiguity lurking here.  The state of
affairs described in the first sentence cannot be held to
\emph{explain} the Lorentz covariance of the laws (surely the
claim that Janssen intends).  But one can take the state of
affairs described in the first sentence as \emph{evidence for} the
Lorentz covariance of the laws.\footnote{\emph{Cf.}\ Janssen's own
distinction between `explanatory' and `evidentiary' uses of
``because'' \citep[459]{janssen02}.}  The passage quoted is true
only if one understands it as making such an evidentiary claim.
And as such, it is (essentially) an unexceptionable statement of
Einstein's 1905 reasoning.\footnote{It is also worth noting that a
curious picture of Einstein's pre-Minkowskian work emerges in
\citet{janssen01}. Janssen stresses that ``In Einstein's special
theory of relativity the Lorentz invariance of these different
laws [of matter and fields] is traced to a common origin'' (p.~6)
and that ``Einstein recognized that Lorentz invariance reflects a
new space-time structure'' (p.~9). But Janssen himself
acknowledges that in 1905 Einstein never talks about space-time,
and that his initial reaction to Minkowski's geometrization of his
1905 theory was negative (p.~9). The careful reader of Janssen's
study would be forgiven for thinking that Einstein misunderstood
his own theory in 1905, or at least its real point of departure
from Lorentz's program. Our position, on the other hand, is that
is that Einstein knew pretty well what he was doing in 1905. In
providing a principle theory approach to deriving the Lorentz
transformations, and hence the non-classical behaviour of rods and
clocks, he was re-systematizing, and giving a different emphasis
to, aspects of the work of Lorentz and Poincar\'e, but not
providing a revolutionary new stance. For Einstein himself, the
real revolution in his 1905 \emph{annus mirabilis} was his light
quantum hypothesis, as is well known.  It has been aptly noted by
\citet[272--4]{staley98} that despite the fact that physicists
seldom distinguished between Lorentz's and Einstein's formulation
of the electron theory in the years immediately following 1905,
Einstein did not seek to redress this situation---indeed he even
referred to ``the theory of Lorentz and Einstein'' in 1906 (though
admittedly in somewhat special circumstances).}

We hope to have made it clear why we do not believe that Minkowski
space-time can play the constructive explanatory role that
Balashov and Janssen would have it serve.  What needs to be
stressed is that this conclusion is appropriate not only for those
who adopt an eliminative relationalist stance towards the ontology
of space-time, and not only in the context of theories with fixed,
absolute space-time structure.  As we argued in
\citet{brownpooley01}, even when one's ontology \emph{includes}
substantival space-time structure, the symmetries of the laws
governing material systems are still crucial in such structure
gaining operational chronogeometric significance.  As we wrote
elsewhere:
\begin{quote}
Despite the fact that in GR one is led to attribute an independent
real existence to the metric field, the general relativistic
explanation of length contraction and time dilation is simply the
dynamical one we have urged in the context of special relativity.
\citep[271]{brownpooley01}
\end{quote}

\section*{Acknowledgements}

We are grateful to Michel Janssen and Simon Saunders for
discussion. This paper was composed during Oliver Pooley's tenure
of a British Academy Postdoctoral Fellowship; he gratefully
acknowledges the support of the British Academy.


\end{document}